\documentclass{article}

\usepackage{microtype}
\usepackage{graphicx}
\usepackage{subcaption}
\usepackage{booktabs} % for professional tables
\usepackage{natbib}
\usepackage{xcolor}
\usepackage{commath}
\usepackage{xcolor}
\usepackage{tabularx,lipsum}
\usepackage{booktabs}
\usepackage{makecell}
\usepackage{hyperref}

\usepackage[accepted]{icml2020}

\icmltitlerunning{Ensemble Learning for CME Arrival Time Prediction}
\newcolumntype{R}{>{\raggedright\arraybackslash}X}
\begin{document}

\twocolumn[
\icmltitle{Ensemble Learning for CME Arrival Time Prediction} 


\begin{icmlauthorlist}
\icmlauthor{Khalid A. Alobaid}{goo}
\icmlauthor{Jason T. L. Wang}{too}

\end{icmlauthorlist}

\icmlaffiliation{goo}{Department of Computer Science, New Jersey Institute of Technology, Newark, NJ 07102, USA}
\icmlaffiliation{too}{College of Computing, New Jersey Institute of Technology, Newark, NJ 07102, USA}

\icmlcorrespondingauthor{Jason T. L. Wang}{wangj@njit.edu}

\icmlkeywords{Machine Learning}

\vskip 0.3in
]


\printAffiliationsAndNotice{}  % leave blank if no need to mention equal contribution

\begin{abstract}
The Sun constantly releases radiation and plasma into the heliosphere. 
Sporadically, the Sun launches solar eruptions such as flares and coronal mass ejections (CMEs).   
CMEs carry away a huge amount of mass and magnetic flux with them. 
An Earth-directed CME can cause serious consequences to the human system. 
It can destroy power grids/pipelines, satellites, and communications. 
Therefore, accurately monitoring and predicting CMEs is important 
to minimize damages to the human system. 
In this study we propose an ensemble learning approach, named CMETNet, 
for predicting the arrival time of CMEs from the Sun to the Earth.
We collect and integrate eruptive events from two solar cycles, \#23 and \#24,
from 1996 to 2021 with a total of 363 geoeffective CMEs.
The data used for making predictions include
CME features, solar wind parameters
and CME images obtained 
from the SOHO/LASCO C2 coronagraph.
Our ensemble learning framework comprises 
regression algorithms 
for numerical data analysis
and a convolutional neural network for image processing. 
Experimental results show that CMETNet performs better than existing machine learning methods reported in the literature,
with a Pearson product-moment correlation coefficient of 0.83 and a mean absolute error of 9.75 hours. 
\end{abstract}

\section{Introduction} 
\label{sec:intro}
\begin{figure}
\begin{center}
\includegraphics[width=7cm]{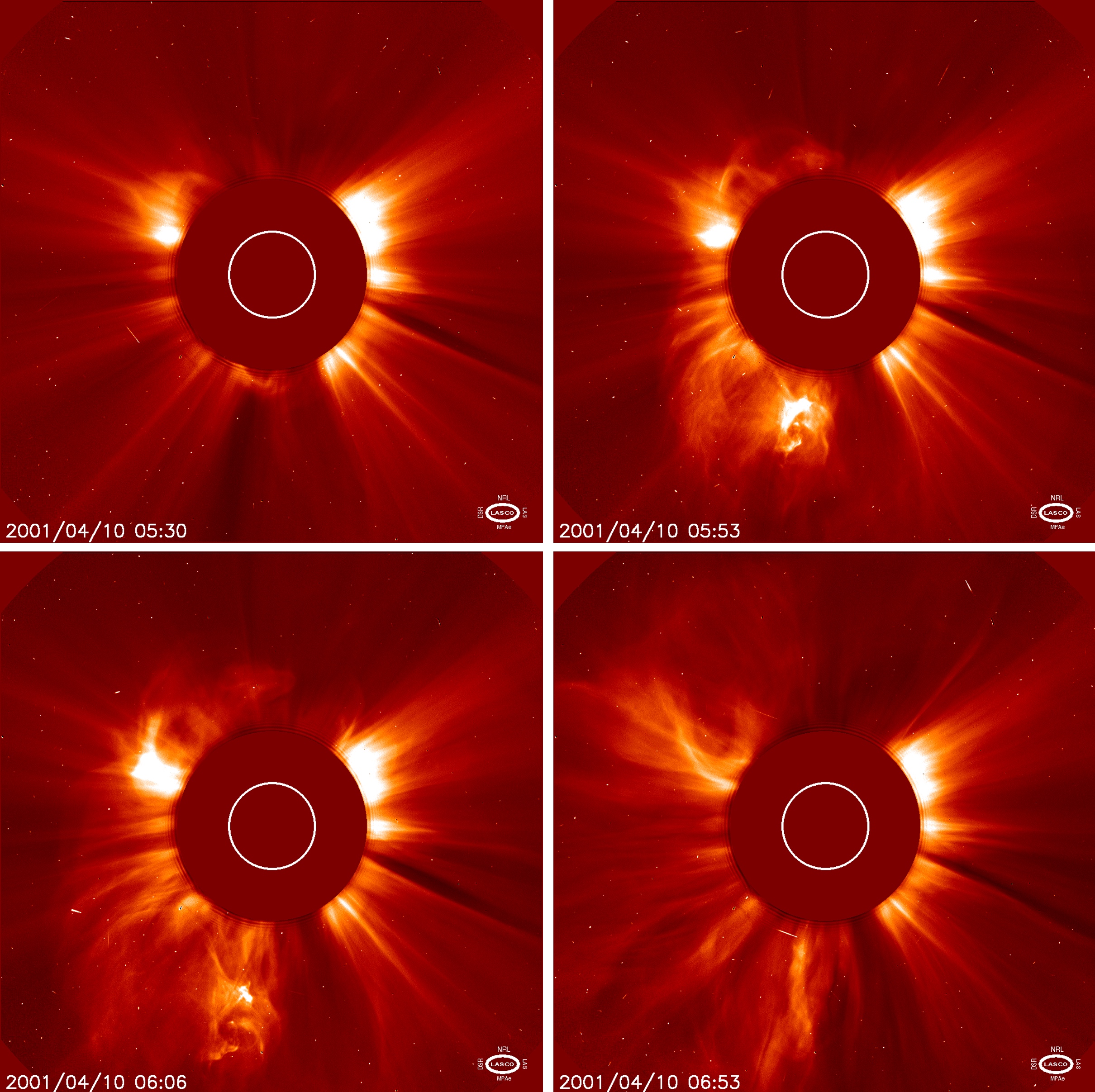}
\end{center}
\caption{Time evolution of a CME,
which occurred on April 10, 2001.
Images were taken from 
\url{https://soho.nascom.nasa.gov}.}
\label{fig:CME-images}
\end{figure}

The launch of the Solar and Heliospheric Observatory (SOHO) mission 
by the European Space Agency (ESA) 
and NASA 
has given scientists opportunities 
to capture full pictures of coronal mass ejections \citep[CMEs;][]{2016GSL.....3....8G}. 
CMEs are the most violent and energetic phenomena that occur within our solar system 
\citep{2005AnGeo..23.1033S,2010cosp...38.1867M, 2012PhDT........22M}.
They are billows of the Sun-based plasma alongside electromagnetic radiation 
ejected out of the solar photosphere in eruptions of energy 
that proliferate in the interplanetary environment. 
CMEs have three particular highlights specifically: 
a center which is thick, 
a front edge which is quite remarkable and a pocket which has low electron density. 
The coronagraph images usually reveal the bright dense edges of the CMEs 
\citep{2009EM&P..104..295G}.
Figure \ref{fig:CME-images} shows 
the time evolution of a CME occurring 
on April 10, 2001 as seen by SOHO.

CMEs are often accompanied with solar flares 
\citep{2002A&ARv..10..313P,2004SpWea...2.2004D,CMH-2019,2019ApJ...877..121L,2020SpWea..1802440J,WCT-2020,2021RAA....21..160A,2022ApJ...931..163S},
though the relationship between CMEs and flares is still under active investigation
\citep{yashiro_gopalswamy_2008,2018ApJ...869...99K,2020ApJ...890...12L,2021MNRAS.506.1916R}.
Everyday magnetically active regions on the surface of the Sun 
undergo various changes to cause CMEs, which in turn travel 
in the interplanetary environment sometimes causing shock waves within 
and are an important factor for space weather forecasting and other related studies. 
Any examination in regards to the Sun-based peculiarities, including yet not restricted to these shocks, 
CMEs and flares, generally requires information connected with the magnetic field of the Sun, 
their association with one another and the encompassing interplanetary medium. 
Thus, it is important to show and comprehend the complex magnetic field and 
its variety regarding time and space. 
The solar activity encompasses 
critical elements in the space weather studies. 
These studies play significant roles for managing satellite tasks, 
space instrument arrangement and their maintenance as well as climate expectation procedures on Earth. 
In addition, CMEs are sources of various space weather effects
in the near-Earth environment, such as geomagnetic storms.
A significant CME can lead to a large-scale, long-term economic and societal catastrophe. 
\citet{2013EOSTr..94..222S} discussed the potential risks from space weather. 
The author pointed out that if a geomagnetic storm similar to the 
Carrington event hit Earth nowadays, 
it could put up to 40 million Americans at risk of power outages 
that could last from days to months with tremendous social and economic impacts.

To mitigate the damages that may be caused by CMEs, 
a large number of methods have been developed to
predict the arrival time of CMEs 
\citep{2014SpWea..12..448Z,2019SpWea..17.1166C,2019RSPTA.37780096V}. 
Among them, two categories of methods, namely physics-based and machine learning-based, are commonly used.
The WSA-ENLIL+Cone (WEC) model \citep{https://doi.org/10.1029/2003JA010135,2018SpWea..16.1245R}
is a popular physics-based method for predicting 
the arrival time of CMEs \citep{2018SpWea..16.1245R}. 
The WEC model has been implemented and used 
by several organizations including
the Space Weather Prediction Center (SWPC) at
the National Oceanic and Atmospheric Administration (NOAA) and 
the Community Coordinated Modeling Center (CCMC)
at the NASA Goddard Space Flight Center (GSFC).
The WEC model is built by utilizing
parameters derived from CME propagation 
including CME speed, density, location, and propagation direction.

In the category of machine learning methods,
\citet{2018ApJ...855..109L} employed a support vector regression (SVR) algorithm to 
estimate the arrival time of  
182 geoeffective CMEs from 1996 to 2015.  
For each CME, the authors considered 
related solar wind parameters and CME features 
obtained from the CME Catalog
of the Large Angle and Spectrometric Coronagraph (LASCO) on board SOHO.
\citet{2019ApJ...881...15W} designed a convolutional neural network to predict CME arrival time. 
The authors considered observed images of 223 geoeffective CMEs from 1997 to 2017
where the images were collected from SOHO/LASCO.
There were totally 1,122 images with 1-10 images per CME. 

In this study, we focus on using machine learning to predict CME transit time, 
which is the time the body of a CME spends traveling in interplanetary space 
from the Sun to the Earth. 
The arrival time can be calculated by adding the transit time to the
onset time of the CME.
We collect and integrate eruptive events from two solar cycles, \#23 and \#24,
from 1996 to 2021 with a total of 363 geoeffective CMEs.
The data used for making predictions include CME features, 
solar wind parameters and CME images obtained 
from the SOHO/LASCO C2 coronagraph.
Our approach, named CMETNet, is based on an ensemble learning framework,
which comprises 
regression algorithms 
for numeral/tabular data analysis
and a convolutional neural network for image processing. 
Experimental results demonstrate the good performance of CMETNet,
with a Pearson product-moment correlation coefficient of 0.83 and a mean absolute error of 9.75 hours. 

The rest of this paper is organized as follows.
Section \ref{sec:datasources} describes the data used in this study. 
Section \ref{sec:method} depicts CMETNet and explains how it works.
Section \ref{sec:expresults} presents experimental results.
Section \ref{sec:conc} concludes the paper and 
points out some directions for future research.

\section{Data} \label{sec:datasources}

\subsection{Data Sources}

Following \citet{2018ApJ...855..109L}, we adopted four CME lists:
(i) the Richardson and Cane (RC) list
\citep{2010SoPh..264..189R}
available at \url{http://www.srl.caltech.edu/ACE/ASC/DATA/level3/icmetable2.htm}, 
(ii) the full halo CME list
maintained by the University of Science and Technology of China (USTC) \citep{Welcomet65:online} and 
available at \url{http://space.ustc.edu.cn/dreams/fhcmes/index.php},
(iii) the George Mason University (GMU) CME/ICME list
\citep{2017SoPh..292...80H} available at
\url{http://solar.gmu.edu/heliophysics/index.php/GMU\_CME/ICME\_List}, 
and 
(iv) the CME Scoreboard maintained by NASA's Community Coordinated Modeling Center (CCMC) 
and available at \url{https://kauai.ccmc.gsfc.nasa.gov/CMEscoreboard/}. 

We combined and cleaned the CMEs in the four lists based 
on the procedures described in \citet{2018ApJ...855..109L}.
We then combined these CMEs from the four lists with those in the 
catalog presented in \citet{2017SoPh..292...30P}.
This process resulted in 
a set of 363 CMEs from 1996 to 2021. 
For each of the 363 CMEs, we collected 12 related CME features and solar wind parameters.
In addition, for each CME, we downloaded its images 
from the SOHO/LASCO C2 coronagraph 
available at 
\url{https://www.swpc.noaa.gov/products/lasco-coronagraph},
as done in \citet{2019ApJ...881...15W}.
Below, we describe our data integration process in detail.

\subsection{Data Integration}

For each CME list (RC, USTC, GMU, CCMC), 
we wrote a 
tracker
with Python to automatically gather 
the onset/appearance times and arrival times of the CMEs in the list.
Then, we combined the CMEs in the four lists into a single dataset,
removed duplicates and scanned the dataset for events 
that occurred within the same hour.
For the events that occurred within the same hour,
we kept only one event among them.
Based on the criteria described in \citet{2018ApJ...855..109L}, 
we ended up with 216, 24, 38, and 113 CMEs 
from the RC, USTC, GMU and CCMC lists respectively. 
For each collected CME,
we kept a record of two parameters:
the onset/appearance time of the CME and
the arrival time of the CME.
In addition to the four CME lists (RC, USTC, GMU, CCMC), 
we considered the catalog presented in \citet{2017SoPh..292...30P},
which contained 266 CME events. 
We combined
these 266 events with the 
216, 24, 38, and 113 CMEs 
from the RC, USTC, GMU and CCMC lists respectively,
and kept only one event among those that occurred within the same hour
in our dataset.
 
\begin{figure}
\begin{center}
\includegraphics[width=7cm]{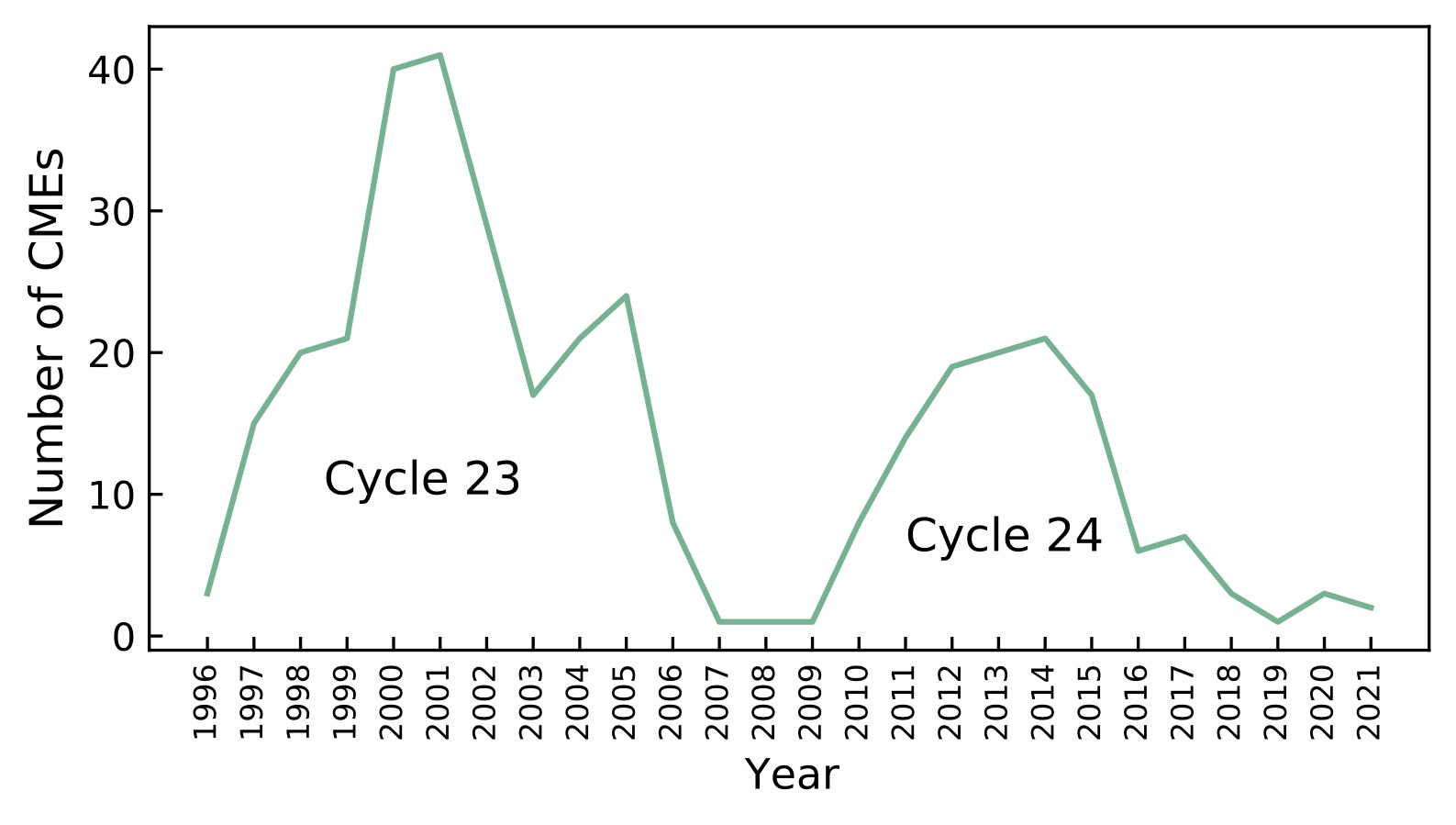}
\end{center}
\caption{Annual counts of the CMEs considered in this study.
These CMEs occurred between August 1996 and May 2021. 
Solar cycle 23 (1996 - 2008) has more CMEs than solar cycle 24 (2008 - late 2019).}
\label{fig:annual_count}
\end{figure}

\begin{figure}
\begin{center}
\includegraphics[width=7cm]{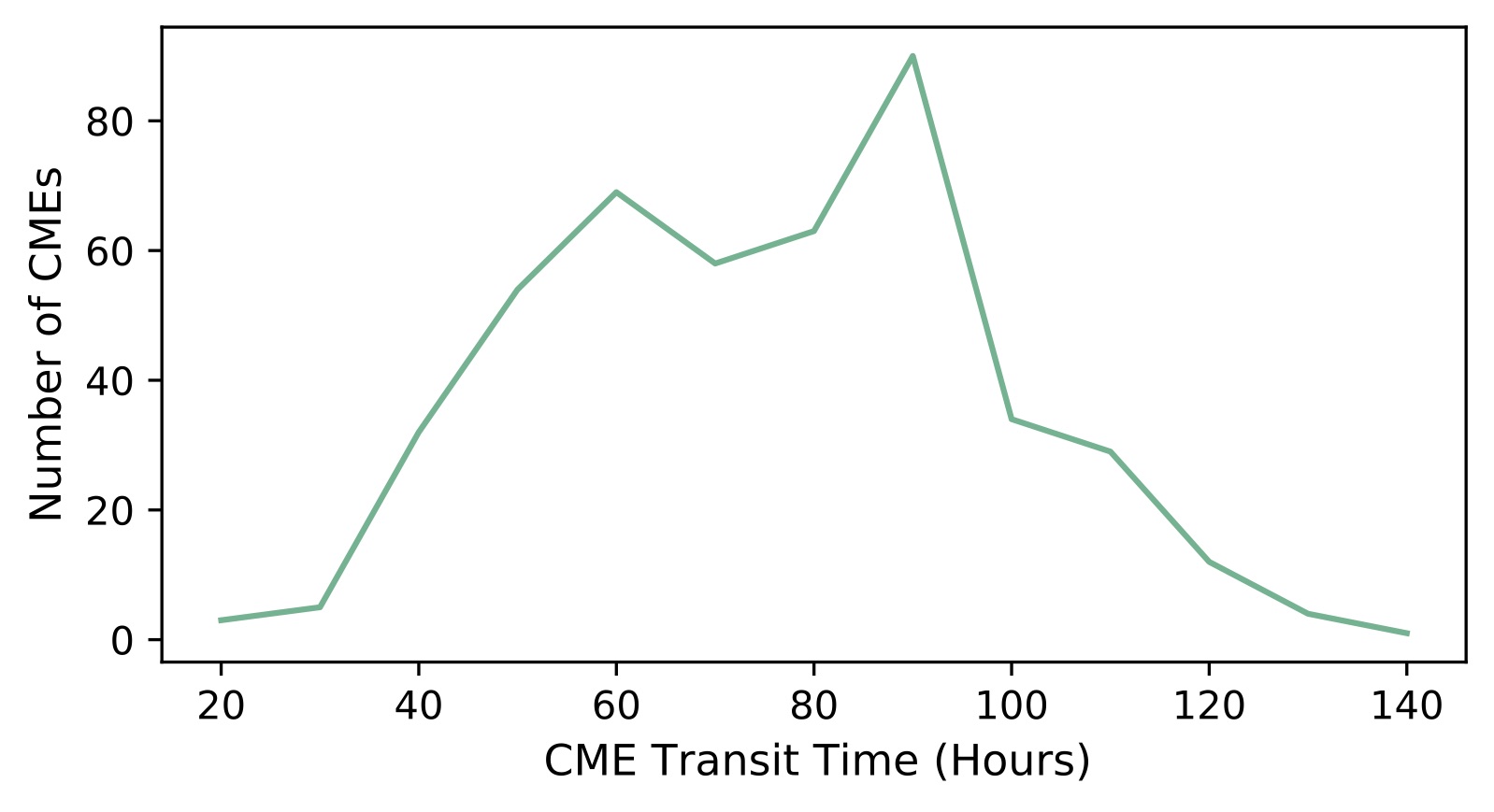}
\end{center}
\caption{Distribution of the 
transit times of the CMEs considered in this study.
For the CMEs, the smallest transit time is 18 hours 
while the largest transit time is 138 hours.}
\label{fig:eventsDistrub}
\end{figure}

This data integration process resulted in 363 geoeffective CMEs.
Figure \ref{fig:annual_count} shows the number of CMEs in each year. 
Solar cycle 23 (1996 - 2008) has 240 CMEs while 
solar cycle 24 (2008 - late 2019) has 118 CMEs.
(There are 5 CMEs between late 2019 and 2021,
totaling 363 CMEs between 1996 and 2021 considered in the study.)
In solar cycle 23, year 2001 has the most CMEs with 41 CMEs in this year.
In solar cycle 24, year 2014 has the most CMEs with 
21 CMEs in this year.
Years 2007, 2008, 2009, and 2019 have the fewest CMEs with only one CME in each of the four years.
It is worth mentioning that the CME peak times 
(i.e., 2001 in solar cycle 23 and 2014 in solar cycle 24)
are consistent with the
sunspot peak times in the two solar cycles \citep{2021MNRAS.506.1916R}. 
Figure \ref{fig:eventsDistrub} shows the 
distribution of CME transit times for the 363 events
in our dataset. 
For the 363 CME events, the smallest transit time is 18 hours 
while the largest transit time is 138 hours.

For each event in our dataset, we obtained its five features from 
the SOHO LASCO CME Catalog 
available at \url{https://cdaw.gsfc.nasa.gov/CME_list/index.html}
and
maintained by
the Coordinated Data Analysis Workshops (CDAW) Data Center at NASA \citep{2009EM&P..104..295G}. 
The five CME features include 
angular width, 
main position angle (MPA), 
linear speed, 
2nd-order speed at final height, 
and mass. 
Specifically, for each event in our dataset,
by using its onset/appearance time as the key, 
we retrieved its five features from 
the LASCO CME Catalog
and merged the features with the arrival time
of the event in our dataset.
For an event $E$ in the dataset whose onset/appearance time does not match 
the onset/appearance time of any CME in the LASCO CME Catalog, 
we selected the five features of the temporally closest matching event in the LASCO CME Catalog that occurred within one hour of $E$ and assigned the five
features to $E$. 
In the LASCO CME Catalog, 
a feature $F$ might have a missing value, and the
missing value was carried over into our dataset.
We employed a data imputation technique by calculating the mean of the available values for the feature $F$
in our dataset
and used the mean to represent the missing value.

To obtain the solar wind parameters of each event in our dataset, 
we followed the approach described in \citet{2018ApJ...855..109L} 
and used the hourly average data extracted from NASA's OMNIWeb at \url{https://omniweb.gsfc.nasa.gov/}. 
We considered seven solar wind parameters:
$B_{x}$, 
$B_{z}$, 
alpha to proton ratio, 
flow longitude, 
plasma pressure, 
flow speed, 
and proton temperature. 
For each event $E$ in our dataset, 
we used $E$'s onset/appearance time $t$ as the key,
retrieved the solar wind parameter values at timestamp $t + 6$
(with hourly average resolution) 
from OMNIWeb,
and assigned the retrieved solar wind parameter values to the 
event $E$.

In addition to the five CME features and seven solar wind parameters,
we downloaded each event's FITS images from the
SOHO/LASCO C2 coronagraph.
We were not able to locate the FITS images for 9
out of the 363 events.
(For these 9 events, we only considered their 
CME features and solar wind parameters when training and testing
our CMETNet framework.)
We followed the approach described in  \citet{2019ApJ...881...15W} 
to match each event in our dataset
with the corresponding FITS files in LASCO.
Specifically, 
for each event in our dataset, we downloaded FITS files
during the period that is between 10 minutes 
before the onset/appearance time of the event and
up to 2 hours after the event. 
The Astropy Python package \citep{2013A&A...558A..33A} 
was used to read each FITS file and produce a CME image as a 2-dimensional (2D) array. 
Image sizes in our dataset are: 
1024 $\times$ 1024 pixels, 
512 $\times$ 512 pixels, or 
256 $\times$ 256 pixels. 
For consistency, we reduced the size of all images 
to 256 $\times$ 256 pixels. 
Images with low quality were discarded. 
In this way, we obtained a total of 2,281 images with 
0-17 images per event.

Finally, we obtained the transit time for a CME event by
calculating the difference in hours
between the onset/appearance time of the event and the arrival time of the event. 
For the 363 geoeffective CMEs in our dataset,
their transit time ranges from 18 hours to approximately 5 days (see Figure \ref{fig:eventsDistrub}).
In other words, the traveling of the CMEs in space from the Sun to the Earth
 takes between 18 hours and approximately 5 days after their first appearance in LASCO. 

\section{Methods} 
\label{sec:method}

\subsection{Overview}
\label{subsec:overview}

\begin{figure*}
\begin{center}
\includegraphics[width=\textwidth]{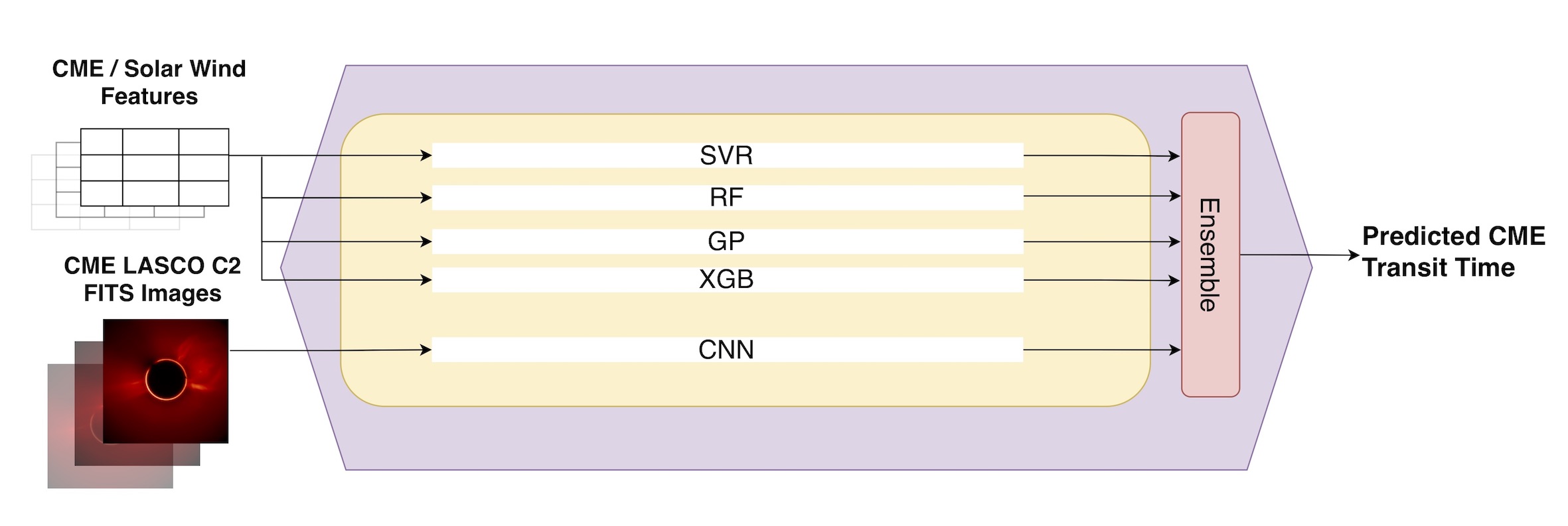}
\end{center}
\caption{Illustration of our CMETNet framework. 
The framework consists of five 
machine learning-based regression models: SVR, RF, GP, XGB and CNN. 
For a CME event $E$,
the first four models (SVR, RF, GP, XGB) accept 
$E$'s CME features and solar wind parameters as input while the
fifth model (CNN) accepts $E$'s CME images as input.
The five regression models are followed by an ensemble method,
which combines the results from the five models and
produces the predicted transit time of $E$.}
\label{fig:model}
\end{figure*}

We adopt an ensemble learning approach to
predicting CME transit time.
This approach uses a set of machine learning models
 whose individual predictions are combined to make a final prediction \citep{DBLP:conf/mcs/Dietterich00}. 
An ensemble method is often more accurate than the individual
machine learning models that form 
the ensemble method \citep{DBLP:conf/mcs/Dietterich00}. 
There are several techniques for constructing an ensemble. 
In this work, we employ bootstrap aggregation, or bagging for short, 
originally proposed by 
\citet{DBLP:journals/ml/Breiman96b},
which works as follows:

\begin{enumerate}
\item[(1)]
Randomly select $N$ training samples with replacement
from a given training set with $M$ samples.
\item[(2)]
Repeat step (1) to generate $L$ training subsets,
$N_{1}$, $N_{2}$, $\ldots$ , $N_{L}$, 
where $L$ is the number of learners forming the ensemble method. 
Notice that the same sample may appear multiple times in the training subsets.
\item[(3)]
Each learner out of the $L$ learners is trained individually
and separately by one of the training subsets, 
$N_{1}$, $N_{2}$, $\ldots$ , $N_{L}$,
with no two learners being trained by the same training subset. 
\item[(4)]
Each trained learner makes a prediction on a given test sample.
The ensemble method produces a final prediction on the test sample through a combination method,
which usually works by taking the average of the $L$ predictions made 
by the $L$ trained learners.
\end{enumerate}

Our ensemble learning framework, named
CMETNet and illustrated in Figure \ref{fig:model},
comprises five base learners:
(i) a support vector regression (SVR) algorithm \citep{cortes1995support},
(ii) a random forest (RF) algorithm \citep{2001MachL..45....5B},
(iii) a Gaussian process (GP) algorithm \citep{YUAN200847},
(iv) an XGBoost (XGB) algorithm \citep{DBLP:conf/kdd/ChenG16}, and
(v) a convolutional neural network (CNN) \citep{DBLP:conf/shape/CunHBB99}.
The first four learners are regression algorithms used for analyzing
the numeral/tabular data related to the CME features and solar wind parameters considered in the study. These four regression algorithms are commonly used in heliophysics and space weather research \citep{2017ApJ...843..104L,2018SpWea..16.1882G,2018ApJ...861..128I,2021ApJS..257...38T,2022ApJS..260...16A}. The fifth learner is a CNN model used for CME image processing.
Below, we describe the five base learners 
and the ensemble method that combines the learners.

\subsection{Base Learners}

Our SVR model is taken from the Sklearn library in Python \citep{DBLP:journals/jmlr/PedregosaVGMTGBPWDVPCBPD11}.
We performed parameter tuning to find the optimal hyperparameters
of the SVR model
by utilizing the GridSearchCV function in Sklearn. 
We used the Radial Basis Function (RBF) kernel
and set the kernel cache size to 200.
Our RF algorithm operates by constructing multiple decision trees during training and 
outputting the mean of the predictions made by the multiple trees during testing. 
We implemented the RF model using Sklearn 
and optimized the model parameters by utilizing Sklearn's RandomizedSearchCV function. 
The number of trees was set to 800 and the number of attributes used to split a node in a tree
was set to the square root of the total number of attributes (which is 12 including
the five CME features and seven solar wind parameters
used in the study).
GP is a non-parametric model that performs data regression using the prior of a Gaussian process. It is a probabilistic approach that provides a level of confidence for the predicted outcome \citep{gortler2019a}. 
We implemented the GP model using Sklearn with two kernels: Whitekernel and RBF. 
A noise of 0.5 was used as a parameter for the Whitekernel. 
XGB is a scalable tree boosting system.
We implemented our XGB model by utilizing the XGBoost package
presented in \citet{DBLP:conf/kdd/ChenG16}. 
The XGBRegressor class was imported into our CMETNet framework from the XGBoost package 
with default parameters. 

\begin{figure*}
\begin{center}
\includegraphics[width=0.8\linewidth]{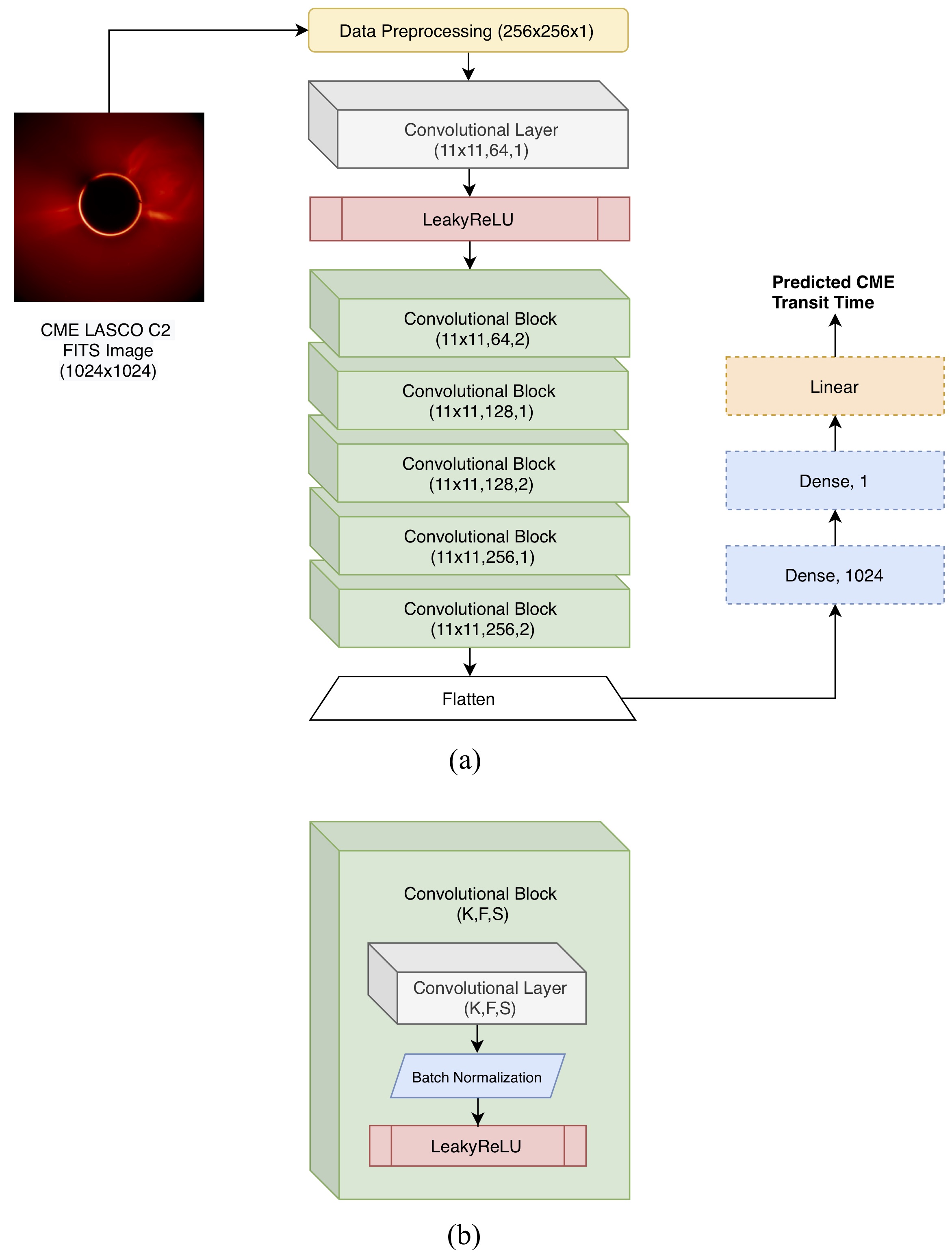}
\end{center}
\caption{Illustration of the CNN model in CMETNet where
the CNN model is used to predict the CME transit time for an input CME LASCO C2 image. 
(a) Overall architecture of the CNN model.
The model starts with a 2D convolutional layer with 64 filters of size $11 \times 11$ and 1 stride,
followed by a Leaky Rectified Linear Unit (LeakyReLU) layer. 
The output of the LeakyReLU layer is then sent to
five convolutional blocks.
The output feature map of the last convolutional block 
is sent to two dense layers with 1024 neurons and 1 neuron, respectively. 
Finally, the model outputs the predicted CME transit time by using a linear activation function. 
(b) Detailed configuration of a convolutional block.
The convolutional block contains 
a 2D convolutional layer with 
F filters where F = 64, 128, 128, 256, 256 respectively and
S strides where S = 2, 1, 2, 1, 2 respectively
and the filter size K = $11 \times 11$.
The 2D convolutional layer is followed by a batch normalization layer, 
which is followed by a LeakyReLU layer.}
\label{fig:CNN_model}
\end{figure*}

Figure \ref{fig:CNN_model} illustrates the architecture of our
CNN model.
This model is used to predict the CME transit time for an input CME LASCO C2 image. 
The model starts with a 2D convolutional layer with 64 filters of size $11 \times 11$ and 1 stride,
followed by a Leaky Rectified Linear Unit (LeakyReLU) layer. 
The output of the LeakyReLU layer is then sent to
five convolutional blocks.
Each convolutional block contains 
a 2D convolutional layer with 
F filters where F = 64, 128, 128, 256, 256 respectively and
S strides where S = 2, 1, 2, 1, 2 respectively
and the filter size K = $11 \times 11$.
The 2D convolutional layer is followed by a batch normalization layer, 
which is followed by a LeakyReLU layer. 
The output feature map of the last convolutional block 
is sent to two dense layers with 1024 neurons and 1 neuron, respectively. 
Finally, the model outputs the transit time by using a linear activation function. 
The regression loss function used by the CNN model is the mean absolute error (MAE) loss function \citep{DBLP:journals/siamrev/Berk92}. 
In training the CNN model, 
we adopt the adaptive moment estimation (Adam) optimizer \citep{Goodfellow-et-al-2016}
to find the optimal parameters/weights of the model.
Please note that the CNN model predicts the transit time from a single CME image
as illustrated in Figure \ref{fig:CNN_model}. 
On the other hand, our CMETNet framework takes all the images associated with a CME event, 
processing and combining the multiple transit times predicted by the CNN model in CMETNet 
using our ensemble learning method as illustrated in Figure \ref{fig:model} to produce the final, predicted transit time of the CME event.
Section \ref{sec:ensemble_learning} below details the 
ensemble learning method.

\subsection{Ensemble Learning}
\label{sec:ensemble_learning}

We adapt the original bagging method described 
in Section \ref{subsec:overview} into the CMETNet framework.
Let $T_{r}$ denote the training set and let $T_{t}$ denote the test set.
Our dataset contains both numerical/tabular data
including CME features and solar wind parameters, which are handled by
the regression algorithms, and CME images, which are handled by the CNN model.
We describe the training procedures for the regression algorithms and
the CNN model separately.

For the regression algorithms, we execute the following steps during training.
\begin{enumerate}
\item[(1)]
Randomly select 200 training samples with replacement 
from $T_{r}$.
Each sample corresponds to a CME event.
\item[(2)]
Repeat step (1) to generate four different training subsets,
$N_{1}$, $N_{2}$, $N_{3}$, $N_{4}$.
Note that the same sample may appear multiple times in
the training subsets.
We also obtain four corresponding validation sets $V_{1}$, $V_{2}$, $V_{3}$, $V_{4}$
where $V_{i} = T_{r} - N_{i}$, $1 \leq i \leq 4$.
Each sample in a validation set also corresponds to a CME event.
\item[(3)]
Each of the base learners/regression algorithms, SVR, RF, GP, XGB, is trained 
individually and separately
by one of the training subsets,
$N_{1}$, $N_{2}$, $N_{3}$, $N_{4}$,
with no two learners being trained by the same training subset.
Each trained learner predicts the transit time
of each sample taken from the corresponding validation set.
The prediction accuracy on the corresponding validation set is measured by
the mean absolute error (MAE)
\citep{DBLP:journals/siamrev/Berk92}.
\item[(4)]
Repeat steps (1) to (3) 100 times to find,
for each base learner,
the best training subset/model 
that yields the highest prediction accuracy on the corresponding validation set.
Save the best model for each of the four base learners.
\end{enumerate}
Since the four base learners SVR, RF, GP, XGB work on 
numerical/tabular data only, 
the data used in the steps (1) to (4) above
for the four base learners
are the CME features and solar wind parameters 
associated with each sample/CME event.

Separately, we train the CNN model in a slightly different way.
Here, each sample corresponds to a CME image rather than a CME event.
A CME event may have multiple images, all of which have the same CME transit time.
Thus, we label these multiple CME images/samples with the same CME transit time.
We then execute the following steps during training.
\begin{enumerate}
\item[(1)] 
Randomly select 10\% of the samples/CME images in the training set $T_{r}$ and
assign the selected samples into the validation set $V$.
\item[(2)]
Use the remaining 90\% training samples in $T_{r}$ along with the 10\% validation samples in $V$ 
to train our CNN model. The validation set $V$ is used to find the optimal parameters and hyperparameters 
to get the best CNN model.
\end{enumerate}

After the training is completed, we obtain the best
SVR, RF, GP, XGB, CNN models. 
During testing, we predict the transit time of each test CME event $E$
in the test set $T_{t}$ based on the following two cases.
\begin{itemize}
    \item $E$ does not have CME images. In this case, $E$ only has numerical/tabular data including CME features and solar wind parameters. Then the predicted transit time for $E$ is the median of the transit times predicted by the four 
    regression models SVR, RF, GP, XGB respectively based on $E$'s CME features and solar wind parameters.
    \item $E$ has CME images, CME features and solar wind parameters.
    Without loss of generality, we assume $E$ has $k$ CME images.
    Then, for each CME image, we use the CNN model to predict the transit time $t_{1}$
    of the image. In addition, we obtain four transit times 
    $t_{2}, t_{3}, t_{4}, t_{5}$
    predicted by
    the four regression models SVR, RF, GP, XGB respectively 
    based on $E$'s CME features and solar wind parameters.
    We take the median of the five predicted transit times
    $t_{1}, t_{2}, t_{3}, t_{4}, t_{5}$.
    $E$ has $k$ CME images, so we obtain $k$ medians.
    Finally, we take the median of the $k$ medians, which 
    is $E$'s predicted transit time produced by our ensemble method (CMETNet).
\end{itemize}

\section{Experiments and Results}
\label{sec:expresults}

\subsection{Experimental Setup and Evaluation Metrics}

The CME Scoreboard maintained by NASA's Community Coordinated Modeling Center (CCMC) 
was launched in 2013.
Our dataset contains 363 CMEs that occurred between 1996 and 2021.
To compare the predictions produced by our CMETNet and those on the CME Scoreboard,
we conducted a 9-fold cross validation experiment as follows.
In each run, data in a year between 2013 and 2021 were used as test data, and all the other data together
were used as training data.
There are 9 years between 2013 and 2021, and hence we had 9 runs/folds.
In each run, the test set and training set were disjoint.
The prediction accuracy in each run was calculated 
and the mean and standard deviation over the 9 runs were plotted.

We adopt two evaluation metrics to quantify the performance of a predictive model.
The first metric is the mean absolute error \citep[MAE;][]{DBLP:journals/siamrev/Berk92}, defined as:
\begin{equation}
\mbox{MAE} = \frac{1}{N} \sum_{i=1}^{N}\left | y_{i} - \hat{y_{i}} \right |
\end{equation}
Here $N$ denotes the number of CMEs in a test set, and
$y_{i}$ ($\hat{y_{i}}$, respectively) denotes the actual (predicted, respectively) transit time of a test CME.
With this metric, we calculate the average of absolute errors between
the actual transit times and predicted transit times
for the CMEs in the test set.
The smaller the MAE is, the more accurate the predictive model is 
and the better performance the model has \citep{Liu_2020_ppmcc}.

The second evaluation metric is the 
Pearson product-moment correlation coefficient \citep[PPMCC;][]{Pearson_PPMCC}, defined as:
\begin{equation}
\mbox{PPMCC} = \frac{E[(X-\mu _X)(Y-\mu _Y)]}{\sigma _X \sigma_Y}
\end{equation}
Here $X$ and $Y$ represent the actual transit times and predicted transit times in the test set;
$\mu_X$ and $\mu_Y$ are the mean of $X$ and $Y$, respectively; 
$\sigma_X$ and $\sigma_Y$ are the standard deviation of $X$ and $Y$ respectively; 
and $E(\cdot)$ is the expectation. 
The value of the PPMCC ranges from $-1$ to 1. 
A value of 1 means that there is a linear equation relationship between $X$ and $Y$, 
where $Y$ increases as $X$ increases. 
A value of $-1$ means that all data points lie on a line for which $Y$ decreases as $X$ increases. 
A value of zero means that there is no linear correlation between the variables $X$
and $Y$ \citep{Liu_2020_ppmcc}.

\subsection{Ablation Studies}

\begin{figure*}[ht]
\begin{center}
\includegraphics[width=0.70\textwidth]{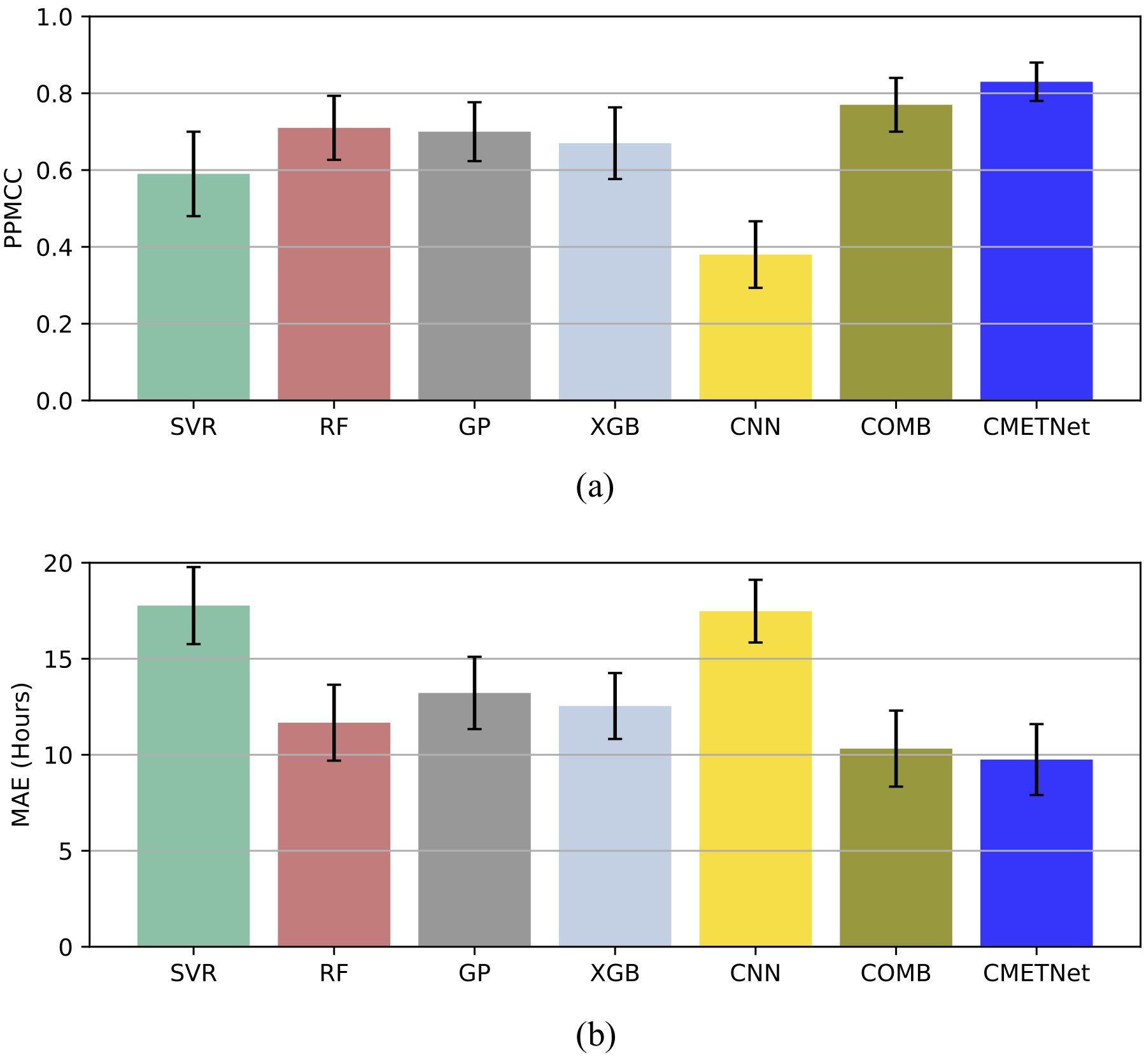}
\end{center}
\caption{Results of the ablation tests for assessing the components 
(SVR, RF, GP, XGB, CNN)
of CMETNet where
COMB is the ensemble model of SVR, RF, GP and XGB.
(a) Pearson product-moment correlation coefficients (PPMCCs) of the tested models.
(b) Mean absolute errors (MAEs) of the tested models.
CMETNet achieves the best performance among all the tested models.}
\label{fig:ablation}
\end{figure*}

In this subsection, we performed ablation tests to analyze and
evaluate the components of our CMETNet framework. 
We considered five subunits from  CMETNet:
SVR, RF, GP, XGB and CNN.
In addition, we considered COMB, which was the ensemble of SVR, RF, GP and XGB.
That is, COMB was obtained by removing CNN from CMETNet.
Figure~\ref{fig:ablation} presents the PPMCC and MAE results of the seven methods: 
SVR, RF, GP, XGB, CNN, COMB, and CMETNet. 
In the figure, 
each colored bar represents the mean over the 9 runs and its associated error bar represents the standard deviation divided by 3 (i.e., the square root of the number of runs) \citep{2022XGBoostSYMHbyIong}.
It can be seen from Figure~\ref{fig:ablation} that CMETNet achieves the best performance among the seven methods. CMETNet yields better results than COMB, indicating the importance of including the CNN model in our framework. 
The results based on PPMCC and MAE are consistent.

Each event $E$ in our dataset has 5 CME features
(denoted by F),
7 solar wind parameters (denoted by W) and
0-17 CME images (denoted by I).
To evaluate the effectiveness of these features, parameters and images,
we performed an additional experiment in which we considered the following seven cases:
\begin{itemize}
    \item $E$ contains all of the CME features, solar wind parameters and CME images (denoted by FWI);
   \item
   $E$ contains only the CME features and solar wind parameters (denoted by FW);
   \item
   $E$ contains only the CME features and CME images (denoted by FI);
   \item
   $E$ contains only the solar wind parameters and CME images (denoted by WI);
   \item
   $E$ contains only the CME features (denoted by F);
   \item
   $E$ contains only the solar wind parameters (denoted by W);
   \item
   $E$ contains only the CME images (denoted by I).
\end{itemize}
In each case, we applied CMETNet to the data at hand.
Notice that 
FWI is equivalent to CMETNet.
In the FW case, 
CMETNet amounts to the aforementioned COMB model.
In the I case, CMETNet amounts to the aforementioned CNN model.

\begin{figure*}[ht]
\begin{center}
\includegraphics[width=0.70\textwidth]{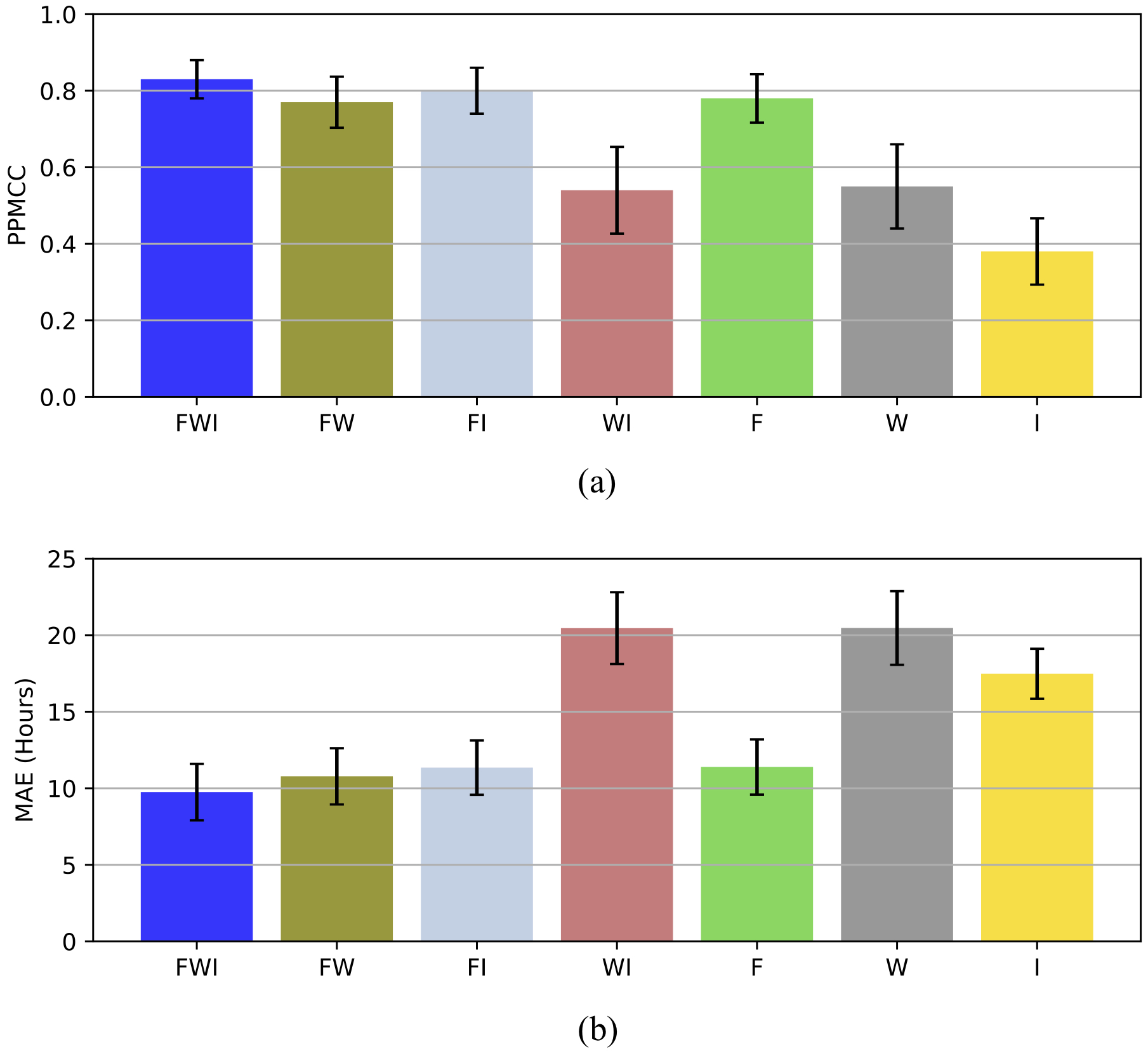}
\end{center}
\caption{Results of the ablation tests for assessing seven cases
(FWI, FW, FI, WI, F, W, I)
where FWI represents the combination of CME features, solar wind parameters and CME images,
FW represents the combination of CME features and solar wind parameters,
FI represents the combination of CME features and CME images,
WI represents the combination of solar wind parameters and CME images,
F represents the CME features,
W represents the solar wind parameters, and 
I represents the CME images.
(a) Pearson product-moment correlation coefficients (PPMCCs) of the tested cases.
(b) Mean absolute errors (MAEs) of the tested cases.
FWI yields the best performance among all the tested cases.}
\label{fig:ablationd}
\end{figure*}

Figure~\ref{fig:ablationd} presents 
the PPMCC and MAE results of the seven cases.
It can be seen from Figure~\ref{fig:ablationd} that
FWI yields the most accurate results with a PPMCC of 0.83 and an MAE of 9.75 hours,
indicating that combining the three types of data (CME features, solar wind parameters and CME images) together
leads to the best performance.
When considering only two types of data together, 
FI has the highest PPMCC of 0.80; 
on the other hand, FW yields the best MAE of 10.32 hours. 
When the three types of data are used individually and separately, 
F yields the best results in terms of both of the two evaluation metrics 
with a PPMCC of 0.78 and an MAE of 11.39 hours. 
Thus, the CME features have higher predictive power than
the solar wind parameters and CME images respectively.

\subsection{Comparison with Related Methods}
In  this  experiment, we compared CMETNet with 
previously published machine learning methods 
for CME arrival time prediction.
\citet{2018ApJ...855..109L} developed an SVR-based method
by utilizing the CME features and solar wind parameters
considered here.
Their method adopted the same SVR model as the one in CMETNet.
We represent their method simply as SVR.
\citet{2019ApJ...881...15W} developed a CNN model 
by utilizing the CME images considered here.
Their CNN model is 
different from the CNN model used in CMETNet.
We represent their CNN model as PCNN (denoting the Previous CNN).
It should be pointed out that, 
although the features/parameters and images 
used by SVR and PCNN respectively are the same as those used 
by CMETNet,
the datasets presented in this work are much larger and more comprehensive than
those presented in the previous studies
\citep{2018ApJ...855..109L,2019ApJ...881...15W}.
As a consequence, the prediction results reported here are not exactly the same
as those given in \citet{2018ApJ...855..109L} and \citet{2019ApJ...881...15W}.

\begin{figure*}[ht]
\begin{center}
\includegraphics[width=0.70\textwidth]{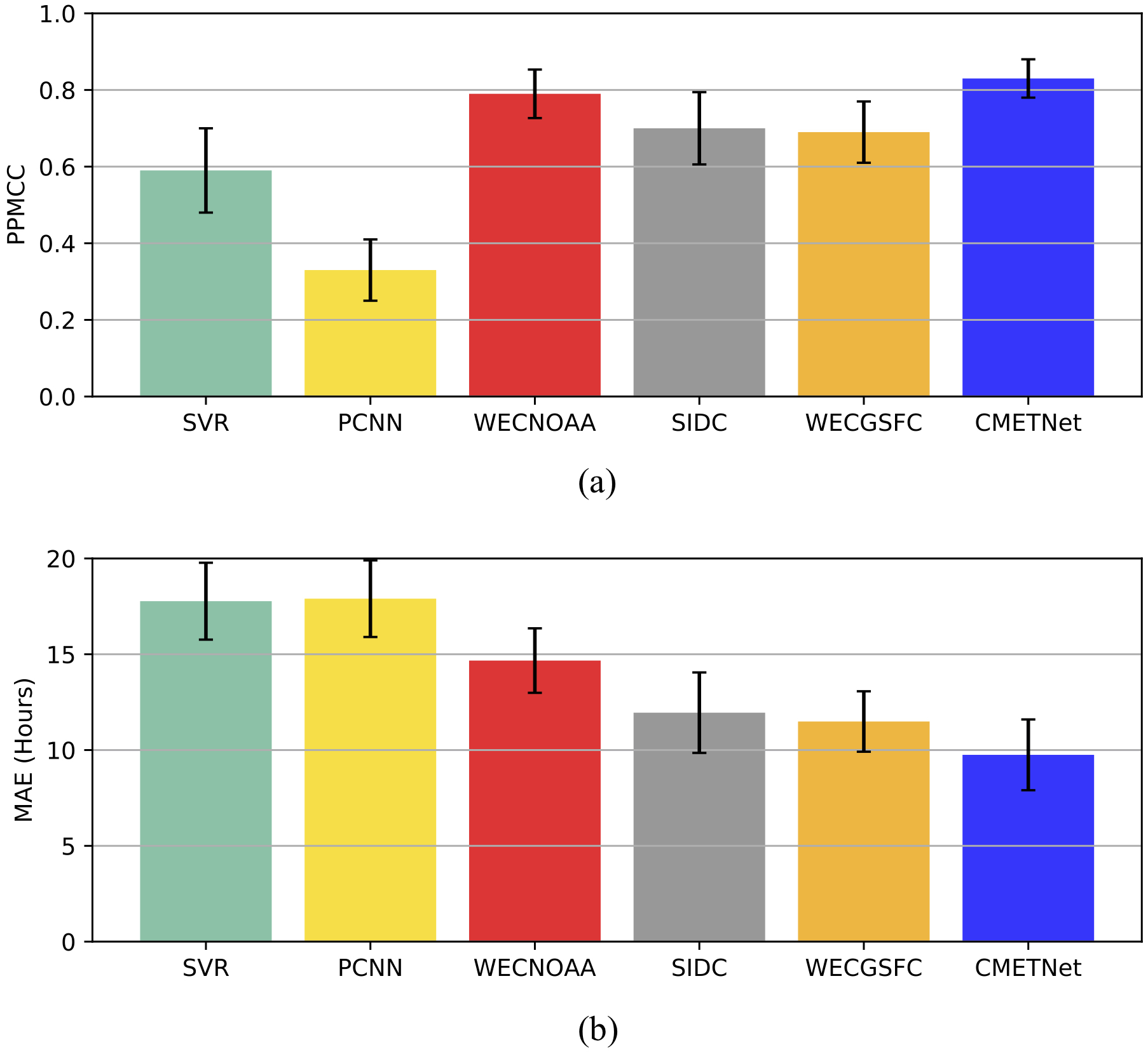}
\end{center}
\caption{Performance comparison of six methods for CME arrival time prediction.
CMETNet is the ensemble learning framework proposed in this paper.
SVR and PCNN are previously published machine learning methods.
WECNOAA, SIDC, and WECGSFC are physics-based models presented on NASA's CCMC CME Scoreboard.
(a) Pearson product-moment correlation coefficients (PPMCCs) of the six methods.
(b) Mean absolute errors (MAEs) of the six methods.
CMETNet outperforms the other five methods in terms of both PPMCC and MAE.}
\label{fig:comp_others}
\end{figure*}

In addition, we compared CMETNet with 
three physics-based models, including 
(i) the WSA-ENLIL+Cone (WEC) model \citep{https://doi.org/10.1029/2003JA010135,2018SpWea..16.1245R} 
implemented by the Space Weather Prediction Center at
the National Oceanic and Atmospheric Administration (NOAA), 
denoted by WECNOAA;
(ii) the model available at 
\url{https://www.sidc.be/}
developed by the 
Solar Influences Data Analysis Center (SIDC) 
at the Royal Observatory of Belgium,
denoted by SIDC; and
(iii) the WSA-ENLIL+Cone (WEC) model 
implemented by the Community Coordinated Modeling Center (CCMC)
at the NASA Goddard Space Flight Center (GSFC),
denoted by WECGSFC.
These three physics-based models 
are considered as the best physical models 
with most accurate prediction results submitted to 
NASA's CCMC CME Scoreboard
at \url{https://kauai.ccmc.gsfc.nasa.gov/CMEscoreboard/}
\citep{2018SpWea..16.1245R}.

Figure~\ref{fig:comp_others} presents the PPMCC and MAE results
of the six methods:
SVR, PCNN, WECNOAA, SIDC, WECGSFC and CMETNet.
Overall, CMETNet yields the most accurate results with a PPMCC of 0.83 and an MAE of 9.75 hours. 
Among the three physics-based methods, 
WECNOAA has the highest PPMCC of 0.79 while WECGSFC has the lowest MAE of 11.49 hours. 
Both CMETNet and the three physics-based methods perform better than
the previously published machine learning methods SVR and PCNN 
\citep{2018ApJ...855..109L,2019ApJ...881...15W}.

\section{Conclusion and Future Work} \label{sec:conc}
In this paper we presented an ensemble framework (CMETNet)
for predicting the arrival time of CME events.
Each event contains 5 CME features
(angular width, 
main position angle, 
linear speed, 
2nd-order speed at final height, 
and mass),
7 solar wind parameters
($B_{x}$, 
$B_{z}$, 
alpha to proton ratio, 
flow longitude, 
plasma pressure, 
flow speed, 
and proton temperature) 
and 0-17 CME images.
The CMETNet framework is composed of five machine learning models including 
support vector regression (SVR), 
random forest (RF),
Gaussian process (GP),
XGBoost (XGB), and
a convolutional neural network (CNN).
Our experimental results demonstrated that
CMETNet outperforms the individual machine learning models
with a PPMCC of 0.83 and an MAE of 9.75 hours.
Furthermore, using all the CME features, 
solar wind parameters and CME images together
yields better performance than using
part of them.

We then compared CMETNet with two previously published machine learning 
methods \citep{2018ApJ...855..109L,2019ApJ...881...15W} 
and three physics-based methods 
(WECNOAA, SIDC, and WECGSFC).
These three physics-based methods
are considered as the best physical models with most accurate prediction results submitted to NASA's
CCMC CME Scoreboard.
Our experimental results showed that
both CMETNet and the three physics-based methods
produce more accurate results than
the two previously published machine learning 
methods, with CMETNet achieving the best performance
among all the methods.

The success of CMETNet is attributed to 
the integration of data from multiple sources and
combination of multiple machine learning models.
Nevertheless, accurately predicting CME arrival time
remains a challenging task as the MAE of CMETNet
(9.75 hours)
is still large, indicating there is much room to improve.
One possible way for improving the performance
of the predictive model is to
combine machine learning with physics-based methods
as done in a recent study \citep{2021EGUGA..23.7661T}.
\citet{2021FrASS...8...58D} reviewed several
physical drag-based model (DBM) tools for predicting
CME arrival time and speed.
In future work we plan to explore the combinations of
these DBM tools and different machine learning models
for more accurate CME arrival time prediction.

Another direction for future research is to 
develop an operational near real-time CME arrival time prediction system
using machine learning.
CME features and solar wind parameters are not applicable here since they are not availabe in real time.
We plan to predict the transit times of CMEs, right after a CME occurred and was caught on LASCO C2 images (\url{https://www.swpc.noaa.gov/products/lasco-coronagraph}) or STEREO COR2 (\url{https://stereo-ssc.nascom.nasa.gov/browse/}), which can be obtained within a few hours.
This near real-time prediction system will require new, advanced CNN models to learn latent features 
or representations from the CME images.
More research is needed to develop such advanced CNN models.

\bibliographystyle{icml2020}

%\bibliography{bibliography}

\begin{thebibliography}{51}
\providecommand{\natexlab}[1]{#1}
\providecommand{\url}[1]{\texttt{#1}}
\expandafter\ifx\csname urlstyle\endcsname\relax
  \providecommand{\doi}[1]{doi: #1}\else
  \providecommand{\doi}{doi: \begingroup \urlstyle{rm}\Url}\fi

\bibitem[{Abduallah} et~al.(2021){Abduallah}, {Wang}, {Nie}, {Liu}, and
  {Wang}]{2021RAA....21..160A}
{Abduallah}, Y., {Wang}, J. T.~L., {Nie}, Y., {Liu}, C., and {Wang}, H.
\newblock {DeepSun: Machine-learning-as-a-service for solar flare prediction}.
\newblock \emph{Research in Astronomy and Astrophysics}, 21\penalty0
  (7):\penalty0 160, August 2021.

\bibitem[{Abduallah} et~al.(2022){Abduallah}, {Jordanova}, {Liu}, {Li}, {Wang},
  and {Wang}]{2022ApJS..260...16A}
{Abduallah}, Y., {Jordanova}, V.~K., {Liu}, H., {Li}, Q., {Wang}, J. T.~L., and
  {Wang}, H.
\newblock Predicting solar energetic particles using {SDO/HMI} vector magnetic
  data products and a bidirectional {LSTM} network.
\newblock \emph{The Astrophysical Journal Supplement}, 260\penalty0
  (1):\penalty0 16, May 2022.

\bibitem[{Astropy Collaboration} et~al.(2013){Astropy Collaboration},
  {Robitaille}, {Tollerud}, {Greenfield}, {Droettboom}, {Bray}, {Aldcroft},
  {Davis}, {Ginsburg}, {Price-Whelan}, {Kerzendorf}, {Conley}, {Crighton},
  {Barbary}, {Muna}, {Ferguson}, {Grollier}, {Parikh}, {Nair}, {Unther},
  {Deil}, {Woillez}, {Conseil}, {Kramer}, {Turner}, {Singer}, {Fox}, {Weaver},
  {Zabalza}, {Edwards}, {Azalee Bostroem}, {Burke}, {Casey}, {Crawford},
  {Dencheva}, {Ely}, {Jenness}, {Labrie}, {Lim}, {Pierfederici}, {Pontzen},
  {Ptak}, {Refsdal}, {Servillat}, and {Streicher}]{2013A&A...558A..33A}
{Astropy Collaboration}, {Robitaille}, T.~P., {Tollerud}, E.~J., {Greenfield},
  P., {Droettboom}, M., {Bray}, E., {Aldcroft}, T., {Davis}, M., {Ginsburg},
  A., {Price-Whelan}, A.~M., {Kerzendorf}, W.~E., {Conley}, A., {Crighton}, N.,
  {Barbary}, K., {Muna}, D., {Ferguson}, H., {Grollier}, F., {Parikh}, M.~M.,
  {Nair}, P.~H., {Unther}, H.~M., {Deil}, C., {Woillez}, J., {Conseil}, S.,
  {Kramer}, R., {Turner}, J. E.~H., {Singer}, L., {Fox}, R., {Weaver}, B.~A.,
  {Zabalza}, V., {Edwards}, Z.~I., {Azalee Bostroem}, K., {Burke}, D.~J.,
  {Casey}, A.~R., {Crawford}, S.~M., {Dencheva}, N., {Ely}, J., {Jenness}, T.,
  {Labrie}, K., {Lim}, P.~L., {Pierfederici}, F., {Pontzen}, A., {Ptak}, A.,
  {Refsdal}, B., {Servillat}, M., and {Streicher}, O.
\newblock {Astropy: A community Python package for astronomy}.
\newblock \emph{Astronomy and Astrophysics}, 558:\penalty0 A33, October 2013.

\bibitem[Berk(1992)]{DBLP:journals/siamrev/Berk92}
Berk, K.~N.
\newblock {Regression analysis (Ashish Sen and Muni Srivastava)}.
\newblock \emph{{SIAM} Rev.}, 34\penalty0 (1):\penalty0 157--158, 1992.

\bibitem[Breiman(1996)]{DBLP:journals/ml/Breiman96b}
Breiman, L.
\newblock Bagging predictors.
\newblock \emph{Machine Learning}, 24\penalty0 (2):\penalty0 123--140, 1996.

\bibitem[{Breiman}(2001)]{2001MachL..45....5B}
{Breiman}, L.
\newblock Random forests.
\newblock \emph{Machine Learning}, 45:\penalty0 5--32, January 2001.

\bibitem[{Camporeale}(2019)]{2019SpWea..17.1166C}
{Camporeale}, E.
\newblock The challenge of machine learning in space weather: Nowcasting and
  forecasting.
\newblock \emph{Space Weather}, 17\penalty0 (8):\penalty0 1166--1207, 2019.

\bibitem[Chen \& Guestrin(2016)Chen and Guestrin]{DBLP:conf/kdd/ChenG16}
Chen, T. and Guestrin, C.
\newblock {XGBoost}: {A} scalable tree boosting system.
\newblock In Krishnapuram, B., Shah, M., Smola, A.~J., Aggarwal, C.~C., Shen,
  D., and Rastogi, R. (eds.), \emph{Proceedings of the 22nd {ACM} {SIGKDD}
  International Conference on Knowledge Discovery and Data Mining}, pp.\
  785--794, 2016.

\bibitem[{Chen} et~al.(2019){Chen}, {Manchester}, {Hero}, {Toth}, {DuFumier},
  {Zhou}, {Wang}, {Zhu}, {Sun}, and {Gombosi}]{CMH-2019}
{Chen}, Y., {Manchester}, W.~B., {Hero}, A.~O., {Toth}, G., {DuFumier}, B.,
  {Zhou}, T., {Wang}, X., {Zhu}, H., {Sun}, Z., and {Gombosi}, T.~I.
\newblock Identifying solar flare precursors using time series of {SDO/HMI}
  images and {SHARP} parameters.
\newblock \emph{Space Weather}, 17\penalty0 (10):\penalty0 1404--1426, 2019.

\bibitem[Cortes \& Vapnik(1995)Cortes and Vapnik]{cortes1995support}
Cortes, C. and Vapnik, V.
\newblock Support-vector networks.
\newblock \emph{Machine Learning}, 20\penalty0 (3):\penalty0 273--297, 1995.

\bibitem[{Daglis} et~al.(2004){Daglis}, {Baker}, {Kappenman}, {Panasyuk}, and
  {Daly}]{2004SpWea...2.2004D}
{Daglis}, I., {Baker}, D., {Kappenman}, J., {Panasyuk}, M., and {Daly}, E.
\newblock {Effects of space weather on technology infrastructure}.
\newblock \emph{Space Weather}, 2\penalty0 (2):\penalty0 S02004, 2004.

\bibitem[Dietterich(2000)]{DBLP:conf/mcs/Dietterich00}
Dietterich, T.~G.
\newblock Ensemble methods in machine learning.
\newblock In Kittler, J. and Roli, F. (eds.), \emph{Multiple Classifier
  Systems, First International Workshop, Proceedings}, volume 1857 of
  \emph{Lecture Notes in Computer Science}, pp.\  1--15. Springer, 2000.

\bibitem[{Dumbovi{\'c}} et~al.(2021){Dumbovi{\'c}}, {{\v{C}}alogovi{\'c}},
  {Martini{\'c}}, {Vr{\v{s}}nak}, {Sudar}, {Temmer}, and
  {Veronig}]{2021FrASS...8...58D}
{Dumbovi{\'c}}, M., {{\v{C}}alogovi{\'c}}, J., {Martini{\'c}}, K.,
  {Vr{\v{s}}nak}, B., {Sudar}, D., {Temmer}, M., and {Veronig}, A.
\newblock {Drag-based model (DBM) tools for forecast of coronal mass ejection
  arrival time and speed}.
\newblock \emph{Frontiers in Astronomy and Space Sciences}, 8:\penalty0 58,
  2021.

\bibitem[Goodfellow et~al.(2016)Goodfellow, Bengio, and
  Courville]{Goodfellow-et-al-2016}
Goodfellow, I., Bengio, Y., and Courville, A.
\newblock \emph{Deep Learning}.
\newblock MIT Press, 2016.

\bibitem[{Gopalswamy}(2016)]{2016GSL.....3....8G}
{Gopalswamy}, N.
\newblock {History and development of coronal mass ejections as a key player in
  solar terrestrial relationship}.
\newblock \emph{Geoscience Letters}, 3:\penalty0 8, 2016.

\bibitem[{Gopalswamy} et~al.(2009){Gopalswamy}, {Yashiro}, {Michalek},
  {Stenborg}, {Vourlidas}, {Freeland}, and {Howard}]{2009EM&P..104..295G}
{Gopalswamy}, N., {Yashiro}, S., {Michalek}, G., {Stenborg}, G., {Vourlidas},
  A., {Freeland}, S., and {Howard}, R.
\newblock {The SOHO/LASCO CME Catalog}.
\newblock \emph{Earth Moon and Planets}, 104\penalty0 (1-4):\penalty0 295--313,
  2009.

\bibitem[{Gruet} et~al.(2018){Gruet}, {Chandorkar}, {Sicard}, and
  {Camporeale}]{2018SpWea..16.1882G}
{Gruet}, M.~A., {Chandorkar}, M., {Sicard}, A., and {Camporeale}, E.
\newblock Multiple-hour-ahead forecast of the {Dst} index using a combination
  of long short-term memory neural network and {Gaussian} process.
\newblock \emph{Space Weather}, 16\penalty0 (11):\penalty0 1882--1896, 2018.

\bibitem[Görtler et~al.(2019)Görtler, Kehlbeck, and Deussen]{gortler2019a}
Görtler, J., Kehlbeck, R., and Deussen, O.
\newblock A visual exploration of {Gaussian} processes.
\newblock \emph{Distill}, 2019.

\bibitem[{Hess} \& {Zhang}(2017){Hess} and {Zhang}]{2017SoPh..292...80H}
{Hess}, P. and {Zhang}, J.
\newblock A study of the {Earth}-affecting {CMEs} of solar cycle 24.
\newblock \emph{Solar Physics}, 292\penalty0 (6):\penalty0 80, 2017.

\bibitem[{Inceoglu} et~al.(2018){Inceoglu}, {Jeppesen}, {Kongstad},
  {Hern{\'a}ndez Marcano}, {Jacobsen}, and {Karoff}]{2018ApJ...861..128I}
{Inceoglu}, F., {Jeppesen}, J.~H., {Kongstad}, P., {Hern{\'a}ndez Marcano},
  N.~J., {Jacobsen}, R.~H., and {Karoff}, C.
\newblock Using machine learning methods to forecast if solar flares will be
  associated with {CMEs} and {SEPs}.
\newblock \emph{The Astrophysical Journal}, 861\penalty0 (2):\penalty0 128,
  2018.

\bibitem[Iong et~al.(2022)Iong, Chen, Toth, Zou, Pulkkinen, Ren, Camporeale,
  and Gombosi]{2022XGBoostSYMHbyIong}
Iong, D., Chen, Y., Toth, G., Zou, S., Pulkkinen, T.~I., Ren, J., Camporeale,
  E., and Gombosi, T. I.~I.
\newblock New findings from explainable {SYM-H} forecasting using gradient
  boosting machines.
\newblock \emph{Earth and Space Science Open Archive}, 2022.

\bibitem[{Jiao} et~al.(2020){Jiao}, {Sun}, {Wang}, {Manchester}, {Gombosi},
  {Hero}, and {Chen}]{2020SpWea..1802440J}
{Jiao}, Z., {Sun}, H., {Wang}, X., {Manchester}, W., {Gombosi}, T., {Hero}, A.,
  and {Chen}, Y.
\newblock Solar flare intensity prediction with machine learning models.
\newblock \emph{Space Weather}, 18\penalty0 (7):\penalty0 e02440, 2020.

\bibitem[{Kawabata} et~al.(2018){Kawabata}, {Iida}, {Doi}, {Akiyama},
  {Yashiro}, and {Shimizu}]{2018ApJ...869...99K}
{Kawabata}, Y., {Iida}, Y., {Doi}, T., {Akiyama}, S., {Yashiro}, S., and
  {Shimizu}, T.
\newblock Statistical relation between solar flares and coronal mass ejections
  with respect to sigmoidal structures in active regions.
\newblock \emph{The Astrophysical Journal}, 869\penalty0 (2):\penalty0 99,
  2018.

\bibitem[LeCun et~al.(1999)LeCun, Haffner, Bottou, and
  Bengio]{DBLP:conf/shape/CunHBB99}
LeCun, Y., Haffner, P., Bottou, L., and Bengio, Y.
\newblock Object recognition with gradient-based learning.
\newblock In Forsyth, D.~A., Mundy, J.~L., Ges{\`{u}}, V.~D., and Cipolla, R.
  (eds.), \emph{Shape, Contour and Grouping in Computer Vision}, volume 1681 of
  \emph{Lecture Notes in Computer Science}, pp.\  319. Springer, 1999.

\bibitem[{Liu} et~al.(2017){Liu}, {Deng}, {Wang}, and
  {Wang}]{2017ApJ...843..104L}
{Liu}, C., {Deng}, N., {Wang}, J. T.~L., and {Wang}, H.
\newblock Predicting solar flares using {SDO/HMI} vector magnetic data products
  and the random forest algorithm.
\newblock \emph{The Astrophysical Journal}, 843\penalty0 (2):\penalty0 104,
  2017.

\bibitem[Liu et~al.(2019)Liu, {Liu}, {Wang}, and {Wang}]{2019ApJ...877..121L}
Liu, H., {Liu}, C., {Wang}, J. T.~L., and {Wang}, H.
\newblock Predicting solar flares using a long short-term memory network.
\newblock \emph{The Astrophysical Journal}, 877\penalty0 (2):\penalty0 121,
  2019.

\bibitem[{Liu} et~al.(2020){Liu}, {Liu}, {Wang}, and
  {Wang}]{2020ApJ...890...12L}
{Liu}, H., {Liu}, C., {Wang}, J. T.~L., and {Wang}, H.
\newblock Predicting coronal mass ejections using {SDO/HMI} vector magnetic
  data products and recurrent neural networks.
\newblock \emph{The Astrophysical Journal}, 890\penalty0 (1):\penalty0 12,
  2020.

\bibitem[Liu et~al.(2020)Liu, Xu, Wang, Jing, Liu, Wang, and
  Wang]{Liu_2020_ppmcc}
Liu, H., Xu, Y., Wang, J., Jing, J., Liu, C., Wang, J. T.~L., and Wang, H.
\newblock Inferring vector magnetic fields from {Stokes} profiles of
  {GST}/{NIRIS} using a convolutional neural network.
\newblock \emph{The Astrophysical Journal}, 894\penalty0 (1):\penalty0 70,
  2020.

\bibitem[{Liu} et~al.(2018){Liu}, {Ye}, {Shen}, {Wang}, and
  {Erd{\'e}lyi}]{2018ApJ...855..109L}
{Liu}, J., {Ye}, Y., {Shen}, C., {Wang}, Y., and {Erd{\'e}lyi}, R.
\newblock A new tool for {CME} arrival time prediction using machine learning
  algorithms: {CAT-PUMA}.
\newblock \emph{The Astrophysical Journal}, 855\penalty0 (2):\penalty0 109,
  2018.

\bibitem[{Maloney}(2012)]{2012PhDT........22M}
{Maloney}, S.
\newblock \emph{{Propagation of Coronal Mass Ejections in the Inner
  Heliosphere}}.
\newblock PhD thesis, School of Physics, Trinity College Dublin, 2012.

\bibitem[{Maloney} et~al.(2010){Maloney}, {Byrne}, {Gallagher}, and
  {McAteer}]{2010cosp...38.1867M}
{Maloney}, S., {Byrne}, J., {Gallagher}, P.~T., and {McAteer}, R.~T.~J.
\newblock {The propagation of a CME front in 3D}.
\newblock In \emph{38th COSPAR Scientific Assembly}, volume~38, pp.\ ~5, 2010.

\bibitem[Odstrcil et~al.(2004)Odstrcil, Riley, and
  Zhao]{https://doi.org/10.1029/2003JA010135}
Odstrcil, D., Riley, P., and Zhao, X.~P.
\newblock Numerical simulation of the 12 {May} 1997 interplanetary {CME} event.
\newblock \emph{Journal of Geophysical Research: Space Physics}, 109\penalty0
  (A2), 2004.

\bibitem[{Paouris} \& {Mavromichalaki}(2017){Paouris} and
  {Mavromichalaki}]{2017SoPh..292...30P}
{Paouris}, E. and {Mavromichalaki}, H.
\newblock Interplanetary coronal mass ejections resulting from {Earth}-directed
  {CMEs} using {SOHO} and {ACE} combined data during solar cycle 23.
\newblock \emph{Solar Physics}, 292\penalty0 (2):\penalty0 30, 2017.

\bibitem[Pearson(1895)]{Pearson_PPMCC}
Pearson, K.
\newblock Note on regression and inheritance in the case of two parents.
\newblock \emph{Proceedings of the Royal Society of London}, 58\penalty0
  (1):\penalty0 240–42, 1895.

\bibitem[Pedregosa et~al.(2011)Pedregosa, Varoquaux, Gramfort, Michel, Thirion,
  Grisel, Blondel, Prettenhofer, Weiss, Dubourg, VanderPlas, Passos,
  Cournapeau, Brucher, Perrot, and
  Duchesnay]{DBLP:journals/jmlr/PedregosaVGMTGBPWDVPCBPD11}
Pedregosa, F., Varoquaux, G., Gramfort, A., Michel, V., Thirion, B., Grisel,
  O., Blondel, M., Prettenhofer, P., Weiss, R., Dubourg, V., VanderPlas, J.,
  Passos, A., Cournapeau, D., Brucher, M., Perrot, M., and Duchesnay, E.
\newblock Scikit-learn: Machine learning in {Python}.
\newblock \emph{J. Mach. Learn. Res.}, 12:\penalty0 2825--2830, 2011.

\bibitem[{Priest} \& {Forbes}(2002){Priest} and {Forbes}]{2002A&ARv..10..313P}
{Priest}, E.~R. and {Forbes}, T.~G.
\newblock {The magnetic nature of solar flares}.
\newblock \emph{Astronomy and Astrophysics Reviews}, 10\penalty0 (4):\penalty0
  313--377, 2002.

\bibitem[{Raheem} et~al.(2021){Raheem}, {Cavus}, {Coban}, {Kinaci}, {Wang}, and
  {Wang}]{2021MNRAS.506.1916R}
{Raheem}, A.-u., {Cavus}, H., {Coban}, G.~C., {Kinaci}, A.~C., {Wang}, H., and
  {Wang}, J. T.~L.
\newblock {An investigation of the causal relationship between sunspot groups
  and coronal mass ejections by determining source active regions}.
\newblock \emph{Monthly Notices of the Royal Astronomical Society},
  506\penalty0 (2):\penalty0 1916--1926, 2021.

\bibitem[{Richardson} \& {Cane}(2010){Richardson} and
  {Cane}]{2010SoPh..264..189R}
{Richardson}, I.~G. and {Cane}, H.~V.
\newblock Near-{Earth} interplanetary coronal mass ejections during solar cycle
  23 (1996 - 2009): Catalog and summary of properties.
\newblock \emph{Solar Physics}, 264\penalty0 (1):\penalty0 189--237, 2010.

\bibitem[{Riley} et~al.(2018){Riley}, {Mays}, {Andries}, {Amerstorfer},
  {Biesecker}, {Delouille}, {Dumbovi{\'c}}, {Feng}, {Henley}, {Linker},
  {M{\"o}stl}, {Nu{\~n}ez}, {Pizzo}, {Temmer}, {Tobiska}, {Verbeke}, {West},
  and {Zhao}]{2018SpWea..16.1245R}
{Riley}, P., {Mays}, M.~L., {Andries}, J., {Amerstorfer}, T., {Biesecker}, D.,
  {Delouille}, V., {Dumbovi{\'c}}, M., {Feng}, X., {Henley}, E., {Linker},
  J.~A., {M{\"o}stl}, C., {Nu{\~n}ez}, M., {Pizzo}, V., {Temmer}, M.,
  {Tobiska}, W.~K., {Verbeke}, C., {West}, M.~J., and {Zhao}, X.
\newblock Forecasting the arrival time of coronal mass ejections: Analysis of
  the {CCMC} {CME Scoreboard}.
\newblock \emph{Space Weather}, 16\penalty0 (9):\penalty0 1245--1260, 2018.

\bibitem[{Schwenn} et~al.(2005){Schwenn}, {dal Lago}, {Huttunen}, and
  {Gonzalez}]{2005AnGeo..23.1033S}
{Schwenn}, R., {dal Lago}, A., {Huttunen}, E., and {Gonzalez}, W.~D.
\newblock {The association of coronal mass ejections with their effects near
  the Earth}.
\newblock \emph{Annales Geophysicae}, 23\penalty0 (3):\penalty0 1033--1059,
  2005.

\bibitem[{Shen} et~al.(2014){Shen}, {Wang}, {Pan}, {Miao}, {Ye}, and
  {Wang}]{Welcomet65:online}
{Shen}, C., {Wang}, Y., {Pan}, Z., {Miao}, B., {Ye}, P., and {Wang}, S.
\newblock {Full-halo coronal mass ejections: Arrival at the Earth}.
\newblock \emph{Journal of Geophysical Research (Space Physics)}, 119\penalty0
  (7):\penalty0 5107--5116, 2014.

\bibitem[{Showstack}(2013)]{2013EOSTr..94..222S}
{Showstack}, R.
\newblock Experts caution about potential increased risks from space weather.
\newblock \emph{EOS Transactions}, 94\penalty0 (25):\penalty0 222--223, 2013.

\bibitem[{Sun} et~al.(2022){Sun}, {Bobra}, {Wang}, {Wang}, {Sun}, {Gombosi},
  {Chen}, and {Hero}]{2022ApJ...931..163S}
{Sun}, Z., {Bobra}, M.~G., {Wang}, X., {Wang}, Y., {Sun}, H., {Gombosi}, T.,
  {Chen}, Y., and {Hero}, A.
\newblock Predicting solar flares using {CNN} and {LSTM} on two solar cycles of
  active region data.
\newblock \emph{The Astrophysical Journal}, 931\penalty0 (2):\penalty0 163,
  2022.

\bibitem[{Tang} et~al.(2021){Tang}, {Zeng}, {Chen}, {Liao}, {Wang}, {Luo},
  {Chen}, {Cui}, {Zhou}, {Deng}, {Li}, {Yuan}, {Hong}, and
  {Wu}]{2021ApJS..257...38T}
{Tang}, R., {Zeng}, X., {Chen}, Z., {Liao}, W., {Wang}, J., {Luo}, B., {Chen},
  Y., {Cui}, Y., {Zhou}, M., {Deng}, X., {Li}, H., {Yuan}, K., {Hong}, S., and
  {Wu}, Z.
\newblock Multiple {CNN} variants and ensemble learning for sunspot group
  classification by magnetic type.
\newblock \emph{The Astrophysical Journal Supplement}, 257\penalty0
  (2):\penalty0 38, 2021.

\bibitem[{Tiwari} et~al.(2021){Tiwari}, {Camporeale}, {Teunissen}, {Foldes},
  {Napoletano}, and {Del Moro}]{2021EGUGA..23.7661T}
{Tiwari}, A., {Camporeale}, E., {Teunissen}, J., {Foldes}, R., {Napoletano},
  G., and {Del Moro}, D.
\newblock {Predicting arrival time for CMEs: Machine learning and ensemble
  methods}.
\newblock In \emph{EGU General Assembly Conference Abstracts}, EGU General
  Assembly Conference Abstracts, pp.\  EGU21--7661, 2021.

\bibitem[{Vourlidas} et~al.(2019){Vourlidas}, {Patsourakos}, and
  {Savani}]{2019RSPTA.37780096V}
{Vourlidas}, A., {Patsourakos}, S., and {Savani}, N.~P.
\newblock {Predicting the geoeffective properties of coronal mass ejections:
  Current status, open issues and path forward}.
\newblock \emph{Philosophical Transactions of the Royal Society of London
  Series A}, 377\penalty0 (2148):\penalty0 20180096, 2019.

\bibitem[{Wang} et~al.(2020){Wang}, {Chen}, {Toth}, {Manchester}, {Gombosi},
  {Hero}, {Jiao}, {Sun}, {Jin}, and {Liu}]{WCT-2020}
{Wang}, X., {Chen}, Y., {Toth}, G., {Manchester}, W.~B., {Gombosi}, T.~I.,
  {Hero}, A.~O., {Jiao}, Z., {Sun}, H., {Jin}, M., and {Liu}, Y.
\newblock Predicting solar flares with machine learning: Investigating solar
  cycle dependence.
\newblock \emph{The Astrophysical Journal}, 895\penalty0 (1):\penalty0 3, 2020.

\bibitem[{Wang} et~al.(2019){Wang}, {Liu}, {Jiang}, and
  {Erd{\'e}lyi}]{2019ApJ...881...15W}
{Wang}, Y., {Liu}, J., {Jiang}, Y., and {Erd{\'e}lyi}, R.
\newblock {CME} arrival time prediction using convolutional neural network.
\newblock \emph{The Astrophysical Journal}, 881\penalty0 (1):\penalty0 15,
  2019.

\bibitem[Yashiro \& Gopalswamy(2008)Yashiro and
  Gopalswamy]{yashiro_gopalswamy_2008}
Yashiro, S. and Gopalswamy, N.
\newblock Statistical relationship between solar flares and coronal mass
  ejections.
\newblock \emph{Proceedings of the International Astronomical Union},
  4\penalty0 (S257):\penalty0 233–243, 2008.

\bibitem[Yuan et~al.(2008)Yuan, Wang, Yu, and Fang]{YUAN200847}
Yuan, J., Wang, K., Yu, T., and Fang, M.
\newblock Reliable multi-objective optimization of high-speed {WEDM} process
  based on {Gaussian} process regression.
\newblock \emph{International Journal of Machine Tools and Manufacture},
  48\penalty0 (1):\penalty0 47--60, 2008.

\bibitem[{Zhao} \& {Dryer}(2014){Zhao} and {Dryer}]{2014SpWea..12..448Z}
{Zhao}, X. and {Dryer}, M.
\newblock {Current status of CME/shock arrival time prediction}.
\newblock \emph{Space Weather}, 12\penalty0 (7):\penalty0 448--469, 2014.

\end{thebibliography}

\end{document}